\documentclass[prb,aps,twocolumn,showpacs]{revtex4}
\usepackage{epsfig}
\begin{document}

\title{Angle-resolved photoemission spectroscopy of electron-doped
cuprate superconductors: Isotropic electron-phonon coupling}

\author{Seung Ryong Park$^1$, D. J. Song$^1$, C. S. Leem$^1$, Chul
Kim$^1$, C. Kim$^{1,*}$, B. J. Kim$^2$, and H. Eisaki$^3$}

\affiliation{$^1$Institute of Physics and Applied Physics, Yonsei
University, Seoul, Korea}

\affiliation{$^2$School of Physics and Center for Strongly
Correlated Materials Research, Seoul National University, Seoul,
Korea}

\affiliation{$^3$AIST Tsukuba Central 2, Umezono, Tsukuba,
Ibaraki, 305-8568, Japan}

\date{\today}

\begin{abstract}
We have performed high resolution angle resolved photoemission
(ARPES) studies on electron doped cuprate superconductors
Sm$_{2-x}$Ce$_x$CuO$_4$ ($x$=0.10, 0.15, 0.18),
Nd$_{2-x}$Ce$_x$CuO$_4$ ($x$=0.15) and Eu$_{2-x}$Ce$_x$CuO$_4$
($x$=0.15). Imaginary parts of the electron removal self energy
show step-like features due to an electron-bosonic mode coupling.
The step-like feature is seen along both nodal and anti-nodal
directions but at energies of 50 and 70 meV, respectively,
independent of the doping and rare earth element. Such energy
scales can be understood as being due to preferential coupling to
half- and full-breathing mode phonons, revealing the phononic
origin of the kink structures. Estimated electron-phonon coupling
constant $\lambda$ from the self energy is roughly independent of
the doping and momentum. The isotropic nature of $\lambda$ is
discussed in comparison with the hole doped case where a strong
anisotropy exists. 
\end{abstract}
\maketitle

Electron-bosonic mode coupling in solids manifests itself as a
slope change in the dispersion or a \textquotedblleft
kink\textquotedblright in angle resolved photoemission (ARPES)
dispersions.\cite{Grimvall} ARPES studies have been extensively
performed to obtain information on the electron-bosonic mode
coupling (EBC)\cite{Lanzara,Johnson,Kaminski,Cuk,Zhou,Kordyuk} in
high temperature superconductors (HTSCs) to find out what mediates
the electron pairing. In spite of the extensive studies,
controversy still exists on what causes the strong kink structure
in ARPES spectra. The controversy can be rooted in the fact that
there exists two bosonic modes with similar energy scales in hole
doped HTSCs that may couple to quasi-particles, i.e., phonons and
magnetic resonance modes. Optical phonons have energies between 40
and 90 meV\cite{Reznik,MeQueeney} while the magnetic resonance
mode from neutron experiments shows an energy scale of 40
meV\cite{Fong,Dai}. Various devised ARPES experiments have been
performed to pin point which mode causes the kink effects.
However, the controversy will continue because the two energy
scales are similar.

The magnetic mode in electron doped HTSCs meanwhile has been
discovered only recently\cite{Wilson,Zhao}. While the energy
scales of the optical phonons in the CuO$_2$ planes are similar
between electron and hole doped cuprates, the magnetic mode in
electron doped cuprates is found to be much smaller ($\sim$ 10 meV
or smaller). Therefore, it should be easier to discern the two
effects in electron doped HTSCs. In addition, recent progress in
the crystal growth technique provides high quality samples
suitable for EBC studies\cite{Park}. This recent progress gives us
an important opportunity to study the kink effects in electron
doped HTSCs. In fact, there is a recent report on this issue from
the scanning tunnelling spectroscopy\cite{Niestemski}. Motivated
by this issue, we performed ARPES experiments on various electron
doped systems of Sm$_{2-x}$Ce$_x$CuO$_4$ (SCCO),
Nd$_{2-x}$Ce$_x$CuO$_4$ (NCCO) and Eu$_{2-x}$Ce$_x$CuO$_4$ (ECCO)
with specific aim on the EBC studies. Our results show a clear
evidence for electron-phonon coupling (EPC). Surprisingly, the
coupling is isotropic in a strong contrast to the hole doped case.
This fact should help to advance our comprehensive understanding
of HTSCs over the entire phase diagram.

SCCO ($x$=0.10, 0.15 and 0.18), NCCO ($x$=0.15) and ECCO
($x$=0.15) single crystals were grown by the travelling-solvent
floating-zone method. Relative Ce concentration was determined by
Ce core-level photoemission and found to be consistent with the
nominal values. NCCO, SCCO and ECCO were reduced by annealing in
N$_2$ for 10 hours at 960, 900, 850 C, respectively, and then in
oxygen for 20 hours at 500 C to induce superconductivity. T$_c$
was determined to be 0, 17, 9, 23 and 0 K by magnetic
susceptibility measurements for SCCO($x$=0.10), SCCO($x$=0.15),
SCCO($x$=0.18), NCCO and ECCO samples, respectively. ARPES
experiments were performed at beamline 5-4 of the Stanford
Synchrotron Radiation Laboratory using 16.5 eV photons with an
energy resolution of 14 meV. Samples were cleaved {\it in situ}
and laser aligned. The chamber pressure was better than $4\times
10^{-11}$ torr and the temperature was kept at 15 K.

\begin{figure}
\centering \epsfxsize=8.3cm \epsfbox{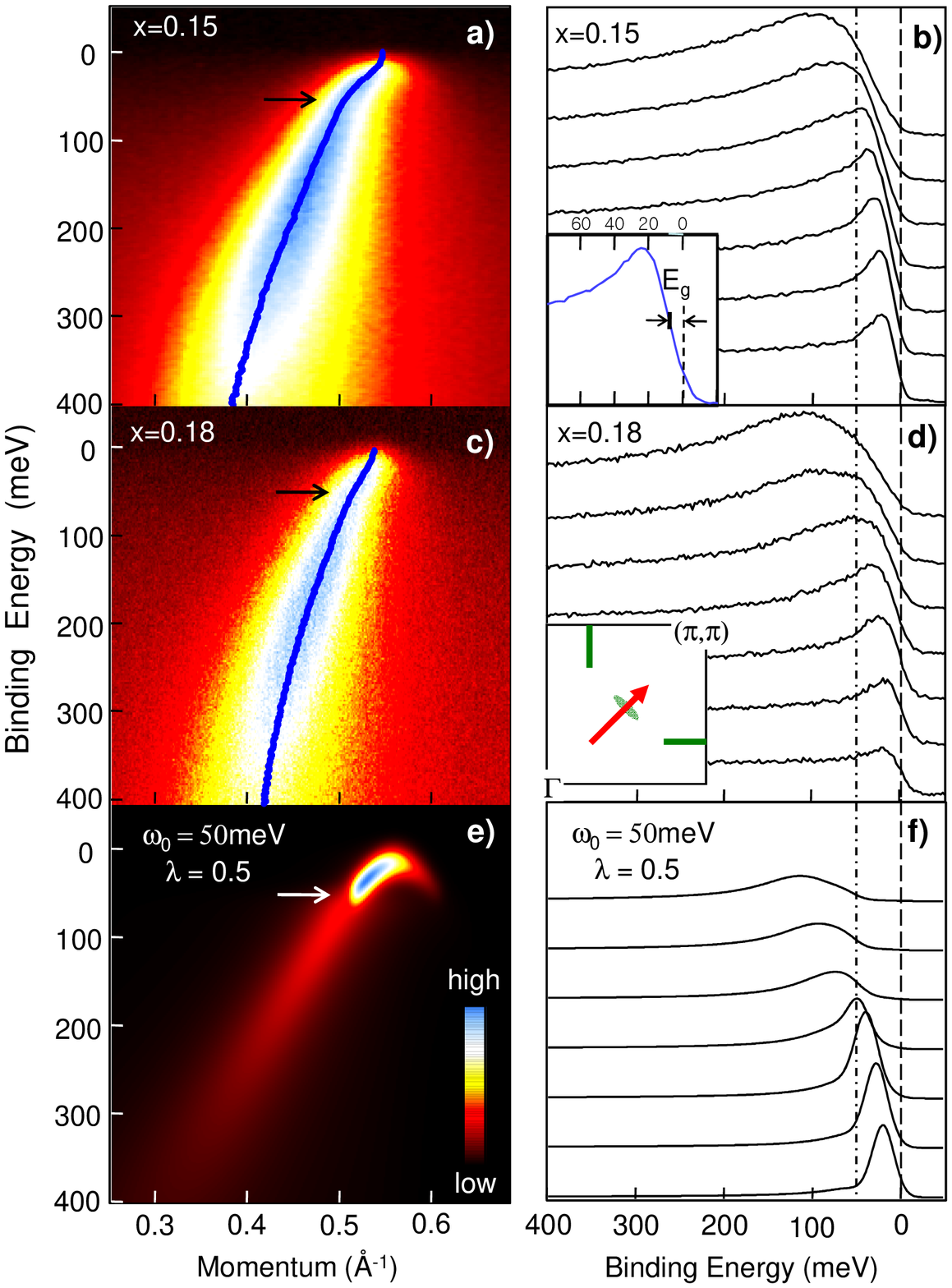} \caption{(Color
online) Intensity maps of ARPES data along the nodal direction
from (a) $x$=0.15 and (c) 0.18 samples of SCCO. The lines in
panels are the MDC dispersions while the arrows mark the energy
where the kink structure appears in the dispersion. Panels (b) and
(d) plot the EDC stacks of the data near the top of the band. The
vertical dash-dot line marks the kink energy. The inset in panel
(b) shows the lowest binding energy feature which remains below
the Fermi level. The inset in panel (d) depicts the direction of
the cuts in the Brillouin zone. Simulation results on EBC for a
folded bare band is shown in panel (e) with EDCs in panel (f) in a
similar fashion to the experimental data.} \label{fig1}
\end{figure}

Fig. 1 (a)-(d) shows ARPES data along the nodal direction from the
SCCO samples with $x$=0.15 and 0.18. $x$=0.1 data is not shown
because the feature is too broad to discuss the EBC. The raw data
and the momentum distribution curve (MDC) dispersions show
kink-like features at around 50 meV (marked by the arrows),
strongly suggesting an EBC at that energy. However, it was shown
that electronic structures of electron doped HTSCs show band
folding effects due to possible $\sqrt{2}\times\sqrt{2}$
ordering\cite{A2,Matsui}. In fact, $x$=0.15 data has a nodal gap,
as shown in the inset of Fig. 1 (b), stemming from a band folding
effect\cite{Park}. In such case, it is difficult to judge if the
bend in the dispersion is a kink or from band folding effects,
especially for the nodal region.

To see if one could still observe a kink structure in spite of a
band folding effect, we plot in Fig. 1 (e) a simulated ARPES
spectral function A(k,$\omega$) of a folded band with a 50 meV
Einstein phonon. The bare band was taken from the published nodal
dispersion of SCCO\cite{Park} and a moderate EPC constant
$\lambda=0.5$ was used. One finds that the kink structure is still
visible for a folded band with a moderate $\lambda$.

We will later provide discussions on the EBC based on the self
energy analysis. However, one can already find some EBC aspects
from the energy distribution curve (EDC) plots. Shown in Fig. 1
(f) are EDCs from the simulation data near the top of the band.
Not only can one still see a kink-like dispersion, but also
observe that the peak becomes considerably broad beyond the mode
energy as scattering is allowed  through EBC. Looking at the
experimental EDCs in Fig. 1 (b) and (d), one can also see that the
peaks quickly broaden beyond the energies marked by the vertical
dash-dot line. Such behavior already suggests the existence of EBC
in these materials.

\begin{figure}
\centering \epsfxsize=8.3cm \epsfbox{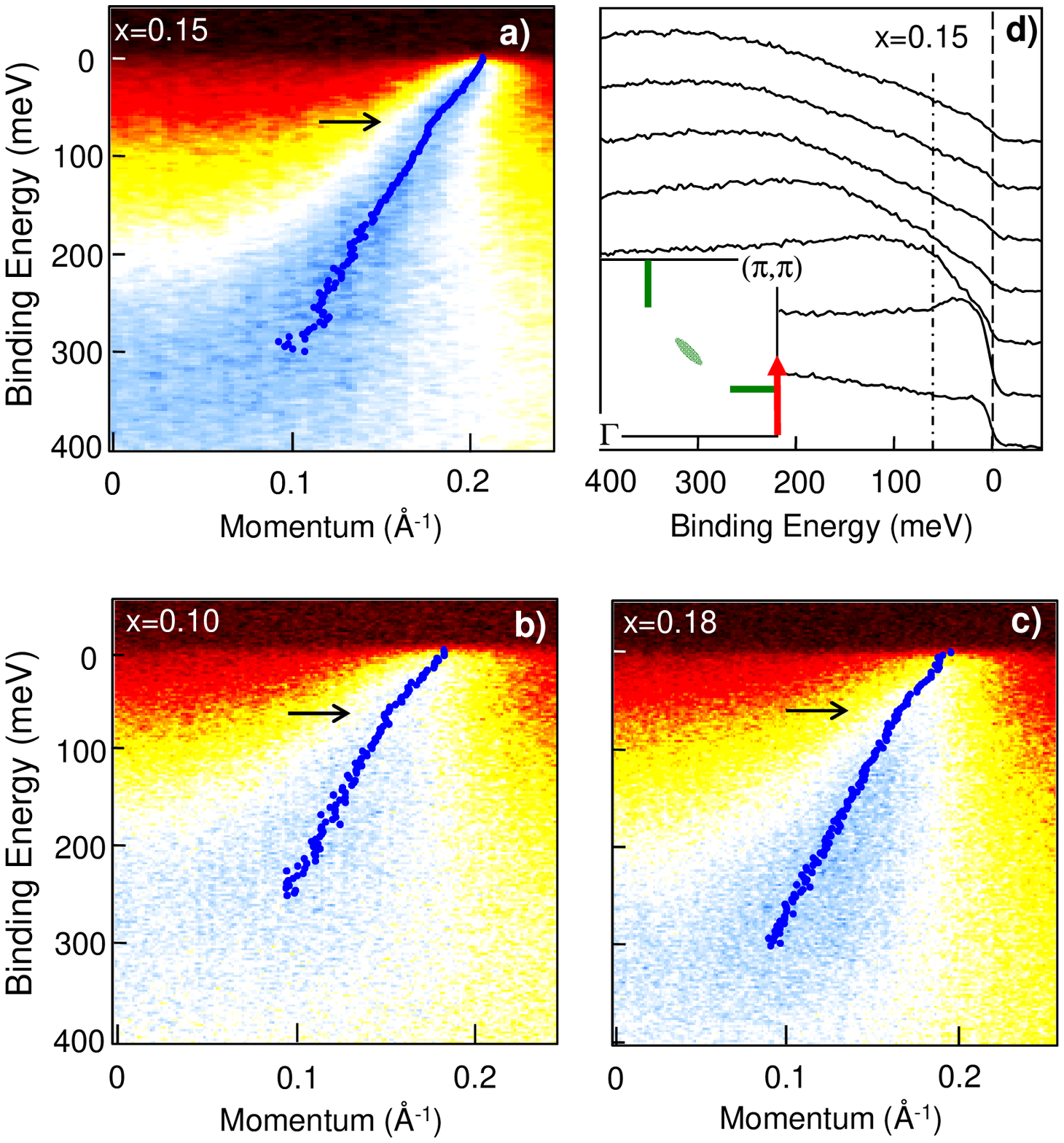} \caption{(Color
online) Intensity maps of ARPES data along the anti-nodal
direction from (a) $x$=0.15, (b) 0.10 and (c) 0.18 samples of
SCCO. As in Fig. 1, the lines are MDC dispersions and the arrows
mark the energy where kinks in the dispersion exist. (d) EDC
stacks of the $x$=0.15 data in (a). The vertical dash-dot line
marks the kink energy.} \label{fig2}
\end{figure}

The kink feature is not only observed along the nodal direction
but also in the anti-nodal direction. Fig. 2 shows ARPES data
along the ($\pi$,0) to ($\pi$,$\pi$) direction. The data show
clear Fermi edges unlike the spectra from the nodal direction.
This is because the near E$_f$ features are from momentum points
that are far enough from the anti-ferromagnetic Brillouin zone
(BZ) boundary and thus the folding effect is minimal. The MDC
dispersions again show kinks at the arrow-marked energy positions.
With minimal folding effects in the region, one can attribute the
kinks to the EBC. EDCs of $x$=0.15 Fig. 2 (d) also show that sharp
peaks at the low binding energy quickly become broad beyond a
certain energy, which is a clear sign of EBC. It is noted that the
characteristic energy where the kink exists is at around 70 meV
(compared to 50 meV for the nodal direction) and is more or less
doping independent.

A natural and important question is whether such kink-like feature
is an intrinsic property of electron doped HTSCs or is specific to
SCCO. In that regards, investigation of other electron doped
compounds is important. Shown in Fig. 3 are the data from
optimally doped ($x$=0.15) NCCO and ECCO. The NCCO nodal cut in
Fig. 3a shows a dispersion that crosses the Fermi energy, unlike
the SCCO with $x$=0.15 case. The EDCs of the data presented in
panel (b) again show sudden broadening beyond a characteristic
energy. On the other hand, ECCO nodal dispersion has a gap of
$\approx 40 meV$ and spectral features are broad. Therefore, for
the same reason as the $x$=0.10 SCCO case, ECCO nodal data is not
presented. This trend of the nodal gap increase when Nd is
replaced by Sm and Eu is consistent with earlier
observation\cite{Ikeda}. For anti-nodal cuts, depicted in Fig. 3
(c) and (d), the dispersive features cross the Fermi level and
show clear kink-like features near 70 meV as was the case for
SCCO. The above results tell us that EBC is present in all the
electron doped HTSCs, independent of the rare earth element.

\begin{figure}
\centering \epsfxsize=8.3cm \epsfbox{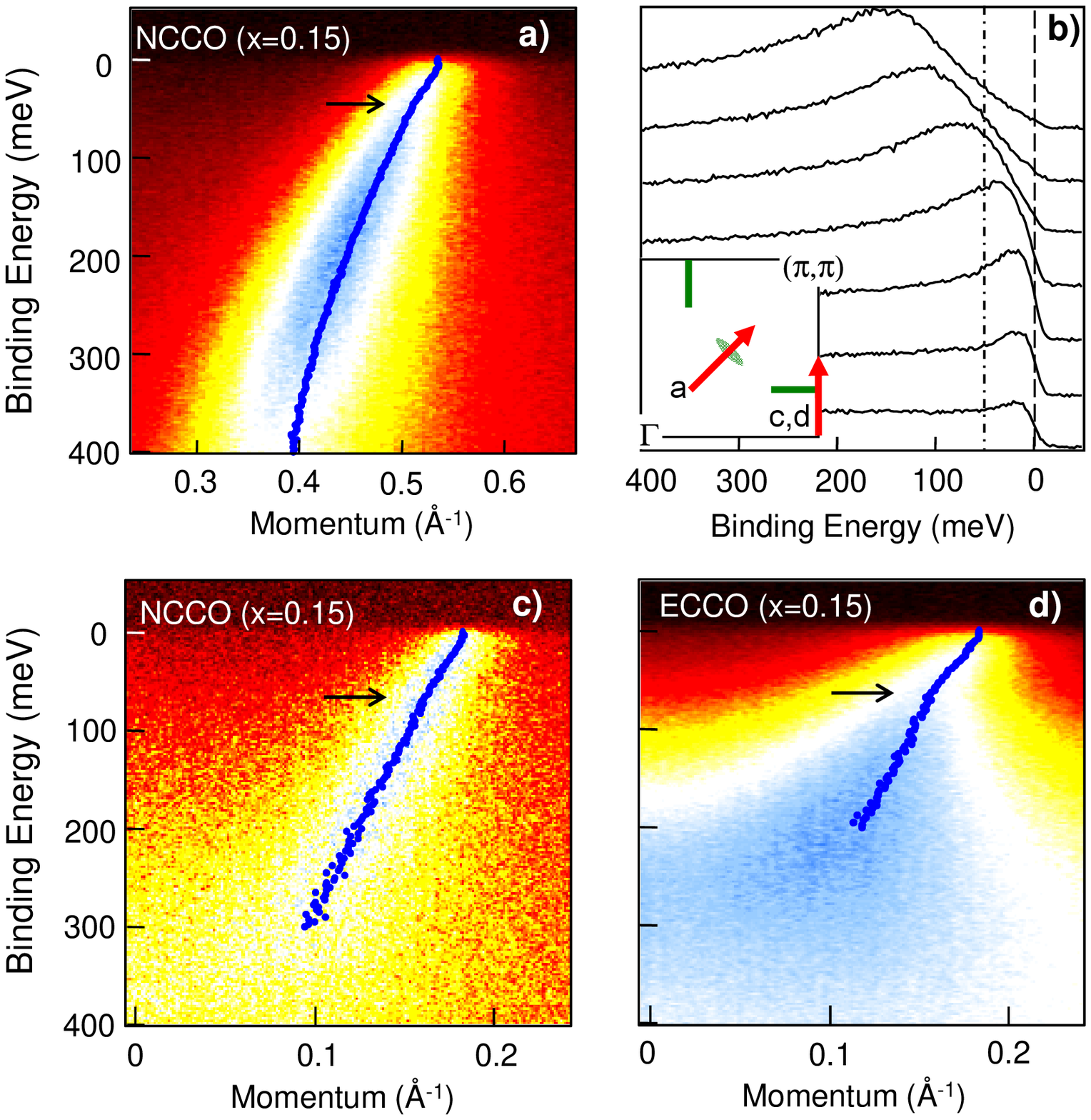} \caption{(Color
online) Intensity maps of ARPES data. (a) Nodal cut for NCCO
($x$=0.15) and (b) EDCs. Anti-nodal cuts for (c) NCCO and (d) ECCO
($x$=0.15). } \label{fig3}
\end{figure}

For a quantitative analysis on EBC, it is useful to look at the
self energies of the spectral functions. Conventionally, the real
part of self energy is obtained by subtracting the bare band
dispersion from the MDC dispersion and imaginary by multiplying
the MDC width with the Fermi velocity. This method, however, is
not applicable to the data from these materials, especially for
the nodal data, because the bare band dispersion is not known due
to a folding effect\cite{Park,Matsui}. One could think about
fitting the ECDs, but it is too unreliable due to the highly
asymmetric lineshape.

In such cases, one may use a newly developed method for
determining the electron $\it removal$ self energy\cite{Chul}.
ARPES intensity I($k$,$\omega$) is proportional to the imaginary
part of electron removal Green's function or the spectral function
A($k$,$\omega$). One could recover the full Green's function if
one knew A($k$,$\omega$) over the entire energy range because the
real and imaginary parts of the Green's function are related
through the Hilbert transform. However, this is not possible
because ARPES measures the spectral function only below the Fermi
energy ($\omega$=0). In that case, one can define electron
$removal$ Green's function which bares information on the electron
dynamics only below $\omega$=0. The rest comes naturally and
one can obtain the imaginary part of the electron removal self
energy Im$\Sigma^R$ without any {\it a priori} assumptions.
Im$\Sigma^R$ is then a simple algebraic expression of the ARPES
spectral function and its Hilbert transform\cite{Chul}. The
advantage of the method is that it can be applied to any line
shape and represents the \textquotedblleft peak
width\textquotedblright of a spectral function. For the purpose of
the discussions given here, Im$\Sigma^R$ can be regarded as
Im$\Sigma$\cite{Chul}.

Fig. 4a shows Im$\Sigma$ of the nodal cuts in Figs. 1 and 3. The
data in panel (a) show a considerable slope change around 50 meV
(vertical dashed line). For the data along the anti-nodal
direction in panel (b), the slope change occurs around an
increased energy of 70 meV. While the 70 meV kink observed near
the anti-nodal direction has been observed in earlier studies on
NCCO\cite{Armitage}, the kink at 50 meV along the nodal direction
in electron doped HTSC is seen for the first time\cite{Schmidt}.
The slope change around 20 meV is an artifact due to the finite
energy resolution.

\begin{figure}
\centering \epsfxsize=8.3cm \epsfbox{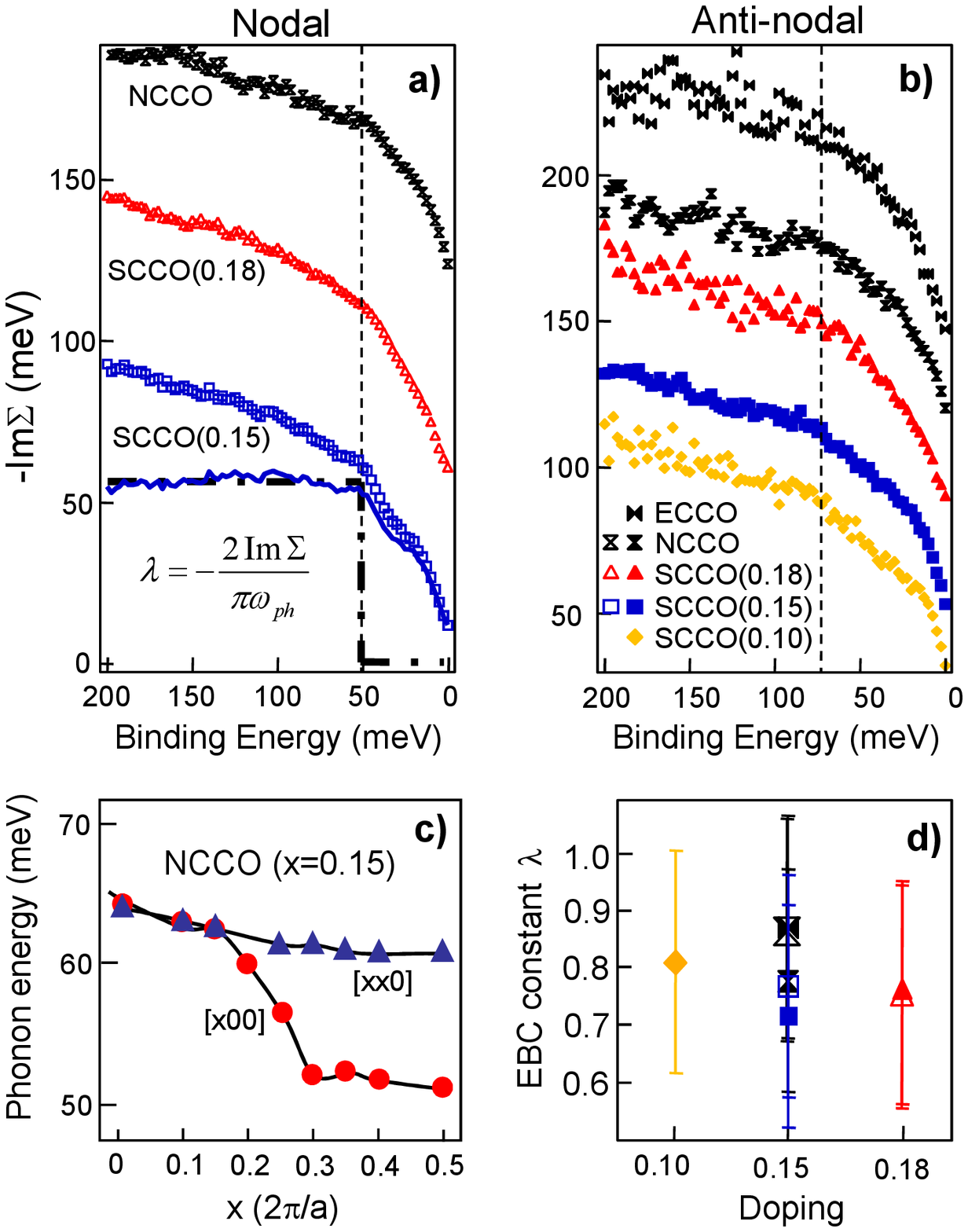} \caption{(Color
online) (a) Imaginary parts of self energies
Im$\Sigma$ of the nodal data in Fig. 1. Curves are offset by 50
meV for clarity. The bottom curves show phonon contribution to
Im$\Sigma$ after a linear background has been subtracted (solid)
and theoretical Im$\Sigma$ with an Einstein phonon (dash-dot).
(b) Im$\Sigma$ for the anti-nodal direction with 30 meV offset.
(c) Experimental phonon dispersions of the bond stretching modes
in NCCO. (d) Estimated EPC constants from Im$\Sigma$ in panels (a)
and (b).} \label{fig4}
\end{figure}

Our data show that there is EBC for both nodal and anti-nodal
directions, at $\approx$50 and $\approx$70 meV, respectively.
These mode energies are roughly independent of doping and rare
earth atoms. Then the question is which bosonic mode causes the
observed kink structures. Recently, it was found through neutron
experiments that the energy of the spin resonance mode in electron
doped HTSCs is at most 10 meV\cite{Wilson,Zhao}. Therefore, the
spin resonance mode can not explain the kinks at 50 and 70 meV in
the ARPES spectra from electron doped HTSCs.

Instead, it can be naturally attributed to the bond stretching
mode of the copper oxygen plane. Fig. 4(c) shows the published
phonon dispersions of bond stretching modes in electron doped
HTSCs\cite{Braden}. We note that the energy range of the phonon
dispersion is similar to that of the kink energy. It is known that
electron coupling to bond stretching phonons is stronger near the
BZ zone boundaries. Phase space argument can be used to show that an
anti-nodal photo-hole couples to full breathing phonons
($q$=[0.5,0.5,0], 60 meV)\cite{Devereaux} and experimental results
show that a nodal photo-hole mostly couples to half breathing mode
($q$=[0.5,0,0], 50 meV)\cite{Cuk,Graf}. This is very much
compatible with the observed energy scales.

As it is clear from the above discussion that the kinks in our
data are from the EPC, we try to estimate the EPC constant.
Conventionally, the EPC constant $\lambda$ is obtained from the
mass re-normalization factor, that is, the ratio between the Fermi
velocities of the bare and experimental dispersions ($\lambda =
v_F^{bare}/v_F^{exp} - 1$). However, this method does not work
when the band dispersion is non-linear, which is the case for
SCCO. On the other hand, one can estimate $\lambda$ from the
Im$\Sigma$ in a more general way. The step height is already a
rough measure of EPC strength. To obtain the step height, one has
to subtract the electron-electron interaction contribution to
Im$\Sigma$. For the energy window in Fig. 4a, Im$\Sigma$ has an
approximate linear background above the kink energy. The solid
blue curve is the resulting Im$\Sigma$ after the linear
background has been subtracted. the resulting Im$\Sigma$ is
compared with an ideal Im$\Sigma$ from a single Einstein phonon at
the kink energy (dash-dot line). Even though the two are different
as there are multiple phonon modes, we use the single Einstein
model for a rough estimate of EPC constant. In that case, the EBC
constant is obtained by a simple equation given in Fig 4a. The
results are plotted in Fig. 4d. Overall, the value is about
$\lambda$=0.8 for both nodal and anti-nodal directions and does
not vary much due to doping and rare earth atom substitution. Even
though the error bars are as large as 40\%, this clearly casts a
sharp contrast with the hole doped cases where EBC constant in the
anti-nodal region is as large as ten times that of the nodal
region\cite{Kaminski}. We can therefore say that EPC in electron
doped cuprates is (relatively) isotropic.

The fact that EPC universally exists in electron doped cuprates
and is similar independent of doping and rare earth elements while
the T$_c$ varies from 25 to 0 K suggests that the observed EPC in
the present work may not play the dominant role in determining the
superconductivity. This fact in turn implies that there must be
other important factors in the superconductivity. The clue to this
question could be found in comparing the present results with that
of hole doped materials for which T$_c$'s are much higher. First
of all, contrary to the electron doped materials, the EBC
$\lambda$ is strongly anisotropic for hole doped cuprates. Such
anisotropy comes from from the fact that a bosonic mode at 40 meV,
either B$_{1g}$ phonon\cite{Cuk,Devereaux} or magnetic resonance
mode\cite{Johnson}, preferentially couples to the electrons in the
anti-nodal region. Electron coupling to this bosonic mode is much
stronger than the coupling to the bond stretching phonon
mode\cite{Kaminski,Devereaux}, and gives strong anisotropic EBC in
hole doped HTSCs\cite{Kaminski}. Secondly, the strong EBC in the
anti-nodal region of hole doped HTSCs is observed only bellow
T$_c$\cite{Johnson} while the kink structure still exists even in
the normal state of electron doped HTSCs as can be seen in ECCO.
Temperature dependence studies on NCCO also show no change in the
kink structure across the T$_c$\cite{Schmidt}

These observations point to the fact that there is a missing
bosonic mode in electron doped HTSCs that couples to the electrons
in the anti-nodal region and causes strong anisotropy in hole
doped materials. This disparity between the two systems may be
related to the fact that the energy position of the van Hove
singularity is at about 350 meV for electron doped while it is
very close to the Fermi energy for hole doped materials.
Independent of the origin, the missing mode could be closely
related to the disparity in T$_c$s between electron- and
hole-doped HTSCs.

We thank T. P. Devereaux, N. P. Armitage and T. Tohyama for
helpful discussions. Experimental supports from D. H. Lu and R. G.
Moore are acknowledged. This work is supported by the KICOS in
K20602000008 and by KOSEF (R01-2006-000-10904-0). SRP acknowledges
support through the BK21 Project and Seoul Science Fellowship.
SSRL is operated by the DOE¡¯s Office of BES.

\end{document}